\documentclass[pmlr]{jmlr}

\usepackage{longtable}

\usepackage{booktabs}
\usepackage[load-configurations=version-1]{siunitx} 
\usepackage{bm}
\usepackage[symbol]{footmisc}

\renewcommand{\thefootnote}{\fnsymbol{footnote}}

\makeatletter
\def\set@curr@file#1{\def\@curr@file{#1}} 
\makeatother

\theorembodyfont{\upshape}
\theoremheaderfont{\scshape}
\theorempostheader{:}
\theoremsep{\newline}

\jmlrvolume{126}
\jmlryear{2021}
\jmlrworkshop{Machine Learning for Healthcare}

\title[MedAug: Contrastive learning leveraging patient metadata]{MedAug: Contrastive learning leveraging patient metadata improves representations for chest X-ray interpretation}

\author{\Name{Yen Nhi Truong Vu$^1$*}
      \Email{ntruongv@stanford.edu}
      \AND
      \Name{Richard Wang$^1$*}
      \Email{rcmwang@stanford.edu}
      \AND
      \Name{Niranjan Balachandar$^2$*}
      \Email{niranja9@stanford.edu}
      \AND
      \Name{Can Liu$^1$}
      \Email{canliu@stanford.edu}
      \AND
      \Name{Andrew Y. Ng$^1$}
      \Email{ang@cs.stanford.edu}
      \AND
      \Name{Pranav Rajpurkar$^1$}
      \Email{pranavsr@cs.stanford.edu}
      \\
      \\
      \normalfont{\textit {*Equal Contribution\\
     $^1$Department of Computer Science, Stanford University\\
     $^2$School of Medicine, Stanford University}}
      } 

\begin{document}

\maketitle

\begin{abstract}
Self-supervised contrastive learning between pairs of multiple views of the same image has been shown to successfully leverage unlabeled data to produce meaningful visual representations for both natural and medical images. However, there has been limited work on determining how to select pairs for medical images, where availability of patient metadata can be leveraged to improve representations.  In this work, we develop a method to select positive pairs coming from views of possibly different images through the use of patient metadata. We compare strategies for selecting positive pairs for chest X-ray interpretation including requiring them to be from the same patient, imaging study or laterality. We evaluate downstream task performance by fine-tuning the linear layer on 1\% of the labeled dataset for pleural effusion classification. Our best performing positive pair selection strategy, which involves using images from the same patient from the same study across all lateralities, achieves a performance increase of 14.4\% in mean AUC from the ImageNet pretrained baseline. Our controlled experiments show that the keys to improving downstream performance on disease classification are (1) using patient metadata to appropriately create positive pairs from different images with the same underlying pathologies, and (2) maximizing the number of different images used in query pairing. In addition, we explore leveraging patient metadata to select hard negative pairs for contrastive learning, but do not find improvement over baselines that do not use metadata. Our method is broadly applicable to medical image interpretation and allows flexibility for incorporating medical insights in choosing pairs for contrastive learning.
\end{abstract}

\section{Introduction}

Self-supervised contrastive learning has recently made significant strides in enabling the learning of meaningful visual representations through unlabeled data \citep{instancedisc, deepinfomax, mocov2, simclrv2}. In medical imaging, previous work has found performance improvement when applying contrastive learning to chest X-ray interpretation \citep{sowrirajan, sriram2021covid, azizi}, dermatology classification (\cite{azizi}) and MRI segmentation \citep{localglobalcontrastive}. Despite the early success of these applications, there is limited work on determining how to improve upon standard contrastive algorithms using medical information \citep{sowrirajan, localglobalcontrastive, textimagecontrastive, clocs}.

In contrastive learning, the selection of pairs controls the information contained in learned representations, as the loss function dictates that representations of positive pairs are pulled together while those of negative pairs are pushed apart \citep{cpc}. For natural images where there are no other types of annotations, positive pairs are created using different augmented views of the same image while negative pairs are views of different images \citep{simclrv2}. \citet{goodview} argue that good positive pairs are those that contain minimal mutual information apart from common downstream task information. In the natural image setting, \citet{tamkin2020viewmaker} train a generative model which learns to produce multiple positive views from a single input.
However, previous contrastive learning studies on medical imaging have not systematically investigated how to leverage patient metadata available in medical imaging datasets to select positive pairs that go beyond 
augmentations of the same image. 

In this work, we propose a method to treat different images that share common properties found in patient metadata as positive pairs in the context of contrastive learning. We demonstrate the application of this method to a chest X-ray interpretation task. Similar to the concurrent work by \citet{azizi}, we experiment with requiring positive pairs to come from the same patient as these images likely share highly similar 
pathological features. However, our method incorporates these positive pairs with possibly different images directly as part of the view generation scheme in a single contrastive pretraining stage, as opposed to \citet{azizi}, which adds a second pretraining stage where a positive pair must be formed by two distinct images. Further, we go beyond the simple strategy of forming a positive pair using any two data points coming from the same patient as in \citet{azizi} and \citet{clocs} and experiment with other metadata such as study number and laterality to identify a pair of images that are likely to have the same pathologies. Although study number has also been leveraged successfully in \citet{sriram2021covid} to create a sequence of pretrained embeddings representating patient disease progression, our work differs in that we use this information specifically to choose positive pairs during the contrastive pretraining stage.

We conduct MoCo-pretraining \citep{mocov2} using these different criteria and evaluate the quality of pretrained representations by freezing the base model and fine-tuning a linear layer using 1\% of the labeled dataset for the task of pleural effusion classification. 
Our contributions are:

\begin{enumerate}
    \item We develop a method, \textit{MedAug}, to use patient metadata to select positive pairs in contrastive learning, and apply this method to chest X-rays for the downstream task of pleural effusion classification. 
    \item Our best pretrained representation achieves a performance increase of 
    14.4\% in mean AUC compared to 
    the ImageNet pretrained baseline, showing that using patient metadata to select positive pairs from multiple images can notably improve representations.
    \item We perform comparative empirical analysis with label information and show that (1) using positive pairs of different images from the same patients that share underlying pathologies improves pretrained representations, and (2) increasing the number of distinct images selected to form positive pairs per image query improves the quality of pretrained representations.  
    \item We perform an exploratory analysis of strategies to select negative pairs using patient metadata, and do not find improvement over the default strategy that does not use metadata.
\end{enumerate}

\begin{figure}[t]
    \floatconts
    {fig:method}
    {\caption{Selecting positive pairs for contrastive learning with patient metadata}}
    {\includegraphics[width=\linewidth]{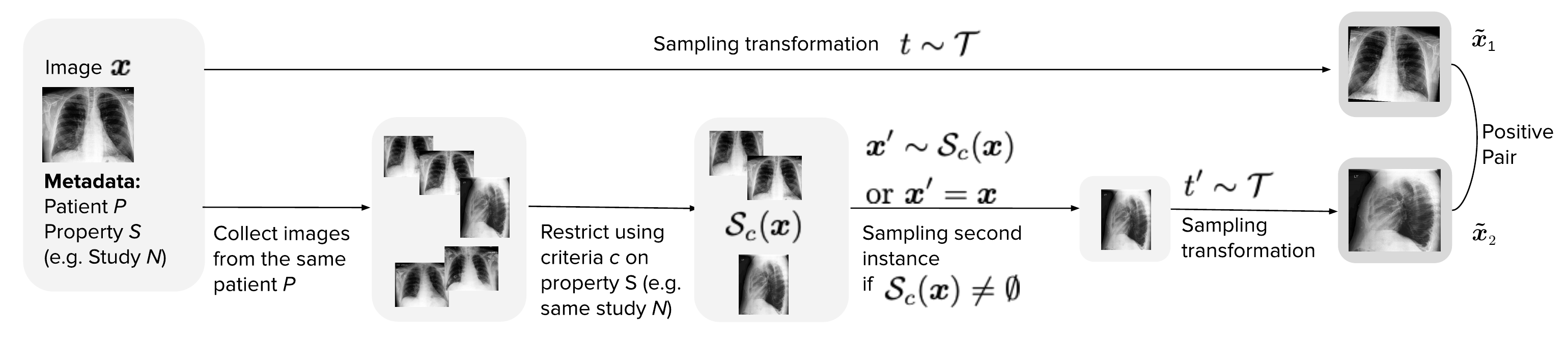}}
\end{figure}

\subsection*{Generalizable Insights about Machine Learning in the Context of Healthcare}

Our work presents methods to improve contrastive pretraining for medical image interpretation. We evaluate our methods on the task of chest X-ray classification, but our approach is generalizable to other medical imaging tasks. First, we demonstrate that leveraging patient metadata -- specifically information about the patient, study, and laterality of the images -- to identify pathological similarity and define positive pairs for contrastive learning improves pretrained representations. This can be applicable to tasks in medical image interpretation where such metadata is available. Second, we show that using two distinct images as positive pairs for contrastive learning provides better pretrained representations than only using two augmentations of the same image. Therefore, other contrastive learning methods for medical imaging may benefit from expanding the construction of positive pairs by using patient metadata to inform this expansion. Third, our analysis indicates that increasing the number of  images available for selecting positive pairs improves pretrained representations. In general, in contrastive learning for medical imaging, using domain-specific information to increase the pool of images to select positive pairs may improve pretrained representations and downstream task performance.

\section{Cohort}
We use CheXpert, a large collection of de-identified chest X-ray images \citep{chexpert}. The dataset consists of 224,316 images from 65,240 patients labeled for the presence or absence of 14 radiological observations. We use these images for pretraining and random samples of 1\% of these images for fine-tuning. The test set consists of 500 additional labeled images from 500 studies not included in the training set.  
We perform fine-tuning experiments on the downstream task of
pleural effusion classification, which was selected based on quality of ground truth labels, clinical importance and prevalence. We evaluate model performance using test set AUC after fine-tuning.

\section{Methods}



\subsection{Selecting positive pairs for contrastive learning with patient metadata}
Given an input image $\bm{x}$, encoder $g$, and a set of augmentations $\mathcal T$, most contrastive learning algorithms involve minimizing the InfoNCE loss
\begin{align} \label{loss}
    \mathcal L(\bm x) = - \log \frac{\exp[g(\bm{\tilde x}_1) \cdot g(\bm{\tilde x}_2)]}{\exp[g(\bm{\tilde x}_1) \cdot g(\bm{\tilde x}_2)] + \sum_{i=1}^K \exp[g(\bm{\tilde x}_1) \cdot g(\bm{z}_i)]}.
\end{align}
Here, the positive pair $(\bm{\tilde x}_1 =t_1(\bm{x}), \bm{\tilde x}_2 =t_2(\bm{x}))$ with $t_1, t_2 \in \mathcal T$ are augmentations of the input image $\bm{x}$, while the negative pairs $(\bm{\tilde x}_1, \bm{z}_i)$, $1 \leq i \leq K$ are pairs of augmentations of different images, with $\bm{z}_i$ coming from either a queue in the case of MoCo or the minibatch in the case of SimCLR. Recognizing that many augmentation strategies available for natural images are not applicable to medical images, \citet{sowrirajan} restrict $\mathcal T$ to be the set of simple augmentations such as horizontal flipping and random rotation between -10 to 10 degrees. 
As a result, their method can be thought of as instance discrimination, as $\bm{\tilde x}_1$ and $\bm{\tilde x}_2$ must come from the same image input.

In this work, we propose MedAug, a method to use multiple images as a way to increase the number of positive pair choices. Beyond the disease labels, we can use patient metadata such as patient number, study number, laterality, patient historical record etc. to create appropriate positive pairs. Formally, we can use patient metadata to obtain an enhanced augmentation set dependent on $\bm x$ as follows
\begin{align}\label{eqn:enhanced_aug}
    \mathcal T_{\text{enhanced}}(\bm{x}) &= \begin{cases}\{ t_i(\bm{x}') | t_i \in \mathcal T, \bm{x}' \in \mathcal S_c(\bm{x})\} & \text{if } \mathcal S_c(\bm x) \neq \emptyset\\ 
    \mathcal T(\bm x) & \text{otherwise}
    \end{cases}
\end{align}
where $\mathcal S_c(\bm x)$ is the set of all images satisfying some predefined criteria $c$ in relation to the properties of $\bm x$. The criteria for using the metadata could be informed by clinical insights about the downstream task of interest.


We apply this method on chest X-ray interpretation and pretrain ResNet-18 models using MoCo v2 \citep{mocov2} with learning rate of $10^{-4}$, batch size of 16, temperature of 0.2, MLP projection, and 20 epochs. 
Since the downstream task is disease classification, we experiment with using $\mathcal S_{\text{same patient}}(\bm x)$ since images from the same patient are likely to share high amount of 
pathological features. We also experiment with further applying criteria on study numbers as well as laterality. An example application of the method is illustrated in Figure \ref{fig:method}. Given an image query, the method collects all images from the same patient as the query, chooses a subset of the collection using the given criteria, then sample an image from the subset.  Finally, an augmentation of the image together with the augmentation of an initial query image form a positive pair. 



\subsection{Fine-tuning and evaluation}
We evaluate the pretrained representations by (1) training a linear classifier on outputs of the frozen encoder using labeled data and (2) end-to-end fine-tuning. Pretrained checkpoints are selected with k-nearest neighbors algorithm \citep{wu2018unsupervised} based on Faiss similarity search and clustering library \citep{JDH17}. To simulate label scarcity encountered in medical contexts, we fine-tune using only 1\% of the labeled dataset. The fine-tuning experiments are repeated on 5 randomly drawn 1\% splits from the labeled dataset to provide an understanding of the model's performance variance. We report the mean AUC and standard deviation over these five 1\% fine-tuning splits. Following 
\cite{chexpert}, we use a learning rate of $ 3 \times 10^{-5}$, batch size of 16 and 95 
epochs for training. 

\section{Experiments}
\subsection{Positive pair selection}
Our formulation of using any set of images $\mathcal S_c(\bm{x})$ from the same patient to enhance the set of augmentations for contrastive learning provides the flexibility of experimenting with different criteria $c$ for constraining $\mathcal S_c(\bm{x})$. We experiment with limiting $\mathcal S_c(\bm{x})$ using properties found in the metadata of the query $\bm{x}$. In particular, we focus on two properties:
\paragraph{Study number.} The study number of an image associated with a particular patient reflects the session in which the image was taken. 
We experiment with three different criteria on study number:
\begin{enumerate}
    \item All studies: no restriction on $\mathcal S_{\text{all studies}}(\bm{x})$ is dependent on the study number of $\bm{x}$
    \item Same study: only images from the same study with $\bm{x}$ belong to $\mathcal S_{\text{same study}}(\bm{x})$
    \item Distinct studies: only images with different study number from $\bm{x}$ belong to $\mathcal S_{\text{distinct studies}}(\bm{x})$
\end{enumerate}
\paragraph{Laterality.} Chest X-rays can be of either frontal (AP/PA) view or lateral view. 
\begin{enumerate}
    \item All lateralities: no restriction on $\mathcal S_{\text{all lateralities}}(\bm{x})$ is dependent on the laterality of $\bm{x}$
    \item Same laterality: only images from the same laterality with $\bm{x}$ belong to $\mathcal S_{\text{same laterality}}(\bm{x})$
    \item Distinct lateralities: only images with a different laterality from that of $\bm{x}$ belongs to $\mathcal S_{\text{distinct lateralities}}(\bm{x})$
\end{enumerate}

\begin{table}[htbp]
\floatconts
{tab:positive}
{\caption{Except for criteria $c$ that involve images from different studies, using images from the same patient to select positive pairs result in improved AUC in downstream pleural effusion classification.}}
{\begin{tabular}{|l|l|l|}
\hline
\textbf{Baseline models} & \textbf{Linear} & \textbf{End-to-end}
\\
\hline \hline 
ImageNet baseline & $0.766 \pm 0.009$ & $0.858 \pm 0.011$ \\
\hline
MoCo v2 baseline \citep{sowrirajan}                                                               & $0.847 \pm 0.007$ & $0.881 \pm 0.017$          \\ \hline
MoCo v2 baseline with random crop scale                                               & $0.864 \pm 0.005$ &   $0.890 \pm 0.026$              \\ \hline 
\hline \textbf{Criteria $c$ for creating} $\mathcal S_c(\bm{x})$ & \textbf{Linear}   & \textbf{End-to-end}            \\ \hline \hline
\begin{tabular}[c]{@{}l@{}}Same patient, same study, same laterality\end{tabular} & $0.862 \pm 0.004$ & $0.894 \pm 0.013$        \\ \hline
Same patient, same study, distinct lateralities   & $0.865 \pm 0.008$ & $0.897 \pm 0.008$          \\ \hline
\begin{tabular}[c]{@{}l@{}}Same patient, same study \end{tabular}    & $0.876 \pm 0.013$ & $0.902 \pm 0.007 $ \\ \hline
\begin{tabular}[c]{@{}l@{}}Same patient, all studies \end{tabular}                   & $0.859 \pm 0.006$ & $0.877 \pm 0.012$          \\ \hline
\begin{tabular}[c]{@{}l@{}}Same patient, distinct studies \end{tabular}                   & $0.848 \pm 0.007$ & $0.874 \pm 0.013$          \\ \hline

\begin{tabular}[c]{@{}l@{}}Same patient, same study with random crop scale   \end{tabular}                           &$\mathbf{0.883 \pm 0.005}$  & $\mathbf{0.906 \pm 0.015}$          \\ \hline
\end{tabular}}
\end{table}

\paragraph{Results.} We report the results of experiments using these criteria in Table \ref{tab:positive}. Except from when $\mathcal S_c(\bm{x})$ includes images with different study numbers from $\bm{x}$, where there is a drop in performance, we see consistent large improvement from the baseline in \cite{sowrirajan}. The best result is obtained when using $\mathcal S_{\text{same study, all lateralities}}(\bm{x})$, the set of images from the same patient and same study as that of $\bm{x}$, regardless of laterality. Incorporating this augmentation strategy while holding other settings from \citet{sowrirajan} constant results in respective gains of 0.029 (3.4\%) and 0.021 (2.4\%) in AUC for the linear model and end-to-end model. We also experiment with including random crop augmentation from MoCo v2 \citep{mocov2}, where the scaling is modified to be $[0.95, 1.0]$ in order to avoid cropping out areas of interest in the lungs. Adding this augmentation to the same patient, same study strategy, we obtain our best pretrained model, which achieves a linear fine-tuning AUC of 0.883 and an end-to-end fine-tuning AUC of 0.906 on the test set, significantly outperforming previous baselines. Results for repeated experiments across other CheXpert competition tasks are included in Appendix \ref{section:new}.

\subsection{Comparative Empirical Analysis}
We perform comparative analysis with labels to understand how different criteria on patient metadata affect downstream performance results seen in Table \ref{tab:positive}.
\subsubsection{All studies  v.s. same study}
We hypothesize that the drop in transfer performance when moving from using images with the same study number to using images regardless of study number is because $\mathcal S_{\text{all studies}}(\bm{x})$ may contain images of a different disease pathology than that seen in $\bm{x}$. As a result, the model is asked to push the representation of a diseased image close to the representation of a non-diseased image, causing poor downstream performance. To test this hypothesis, we carry out an oracle experiment with $\mathcal S_{\text{all studies, same label}}(\bm{x})$, the set of images from the same patient and with the same downstream label as that of $\bm{x}$, regardless of study number. 
\paragraph{Results.} Table \ref{tab:cheat_studies} shows that the model pretrained with $\mathcal S_{\text{all studies, same label}}(\bm{x})$ achieves a respective improvement of 0.034 and 0.022 in AUC over $\mathcal S_{\text{all studies}}(\bm x)$ strategy for the linear model and end-to-end model. This experiment supports our hypothesis that positive pairs from images with different downstream labels hurt performance. 

\begin{table}[htbp]
\floatconts
{tab:cheat_studies}
{\caption{Experiment with and without using downstream labels shows that positive pairs with different labels hurt downstream classification performance.}} 
{\begin{tabular}{|l|l|l|}
\hline
\textbf{Criteria $c$ for creating} $\mathcal S_c(\bm{x})$             & \textbf{Linear}   & \textbf{End-to-end}   \\ \hline \hline
Same patient, all studies                  & $0.859 \pm 0.006$ & $0.877\pm 0.012$ \\ \hline
Same patient, all studies, same disease label as $\bm{x}$ & $\mathbf{0.893 \pm 0.009}$ & $\mathbf{0.899 \pm 0.010}$  \\ \hline
\end{tabular}}
\end{table}
\vspace{-12pt}
\subsubsection{All studies v.s. distinct studies} \label{sec:allvsdis}
There is a further performance drop when moving from using images across all studies of the same patient to images with a different study number from the current query image (Table \ref{tab:positive}). This finding may also support our hypothesis because there is a larger proportion of positive pairs of different disease pathologies in pairs of images from strictly different studies. To make sure this result holds independently of the different number of available images to form pair per query, we repeated these experiments while forcing
$\left|\mathcal S_{\text{same study, all lateralities}}(\bm{x})\right| = \left|\mathcal S_{\text{same study, same laterality}}(\bm{x})\right|$ via random subset pre-selection. Further, we only use distinct images as a pair, i.e. skipping any $\bm x$ with $\mathcal S_c(\bm x) = \emptyset$ in \eqref{eqn:enhanced_aug} in order to remove any possible contribution from positive pairs formed from the same image.

\begin{figure}[h]
    \centering
    \includegraphics[width=.7\linewidth]{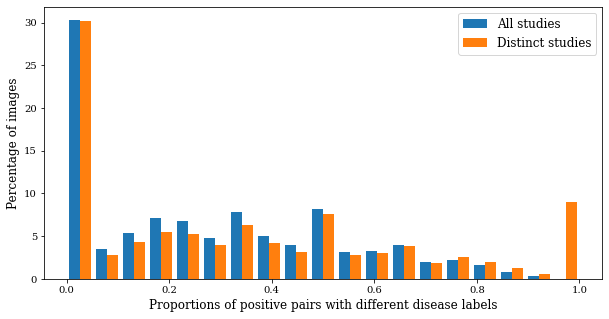}
    \caption{Histogram showing the distribution of the proportions of positive pairs with different disease labels in $\mathcal S_{\text{distinct studies}}(\bm x)$ versus $\mathcal S_{\text{all studies}}(\bm x)$.}
    \label{fig:disease_conflict}
\end{figure}

\paragraph{Results.} Table \ref{tab:control_distinct_studies} shows the same patient, all studies strategy (AUC = 0.848) outperforms the same patient, distinct studies strategy (AUC = 0.792) even when the size of $\mathcal S_c(\bm x)$ is controlled. This supports the hypothesis that a higher proportion of positive pairs with different disease pathologies hurts downstream task performance. 
Downstream performance from the $\mathcal S_{\text{distinct studies}}(\bm x)$ is likely lower than that of $\mathcal S_{\text{all studies}}(\bm x)$ because there is a higher proportion of positive pairs with different disease labels in $\mathcal S_{\text{distinct studies}}(\bm x)$. Figure \ref{fig:disease_conflict} shows that there is almost 9\% of $\bm x$ where $\mathcal S_{\text{distinct studies}}(\bm x)$ contains only images with a different disease label from $\bm x$, whereas this scenario does not appear for $\mathcal S_{\text{all studies}}(\bm x)$.

\begin{table}[htbp]
\floatconts
{tab:control_distinct_studies}
{\caption{Experiments where we force positive pairs to come from different images and control the size of $\mathcal S_c(\bm x)$ shows that higher proportion of pairs with different downstream labels contribute to lower downstream performance.}}
{\begin{tabular}{|l|l|l|}
\hline
\textbf{Criteria $c$ for creating} $\mathcal S(\bm{x})$             & \textbf{Linear}   & \textbf{End-to-end}   \\ \hline \hline
Same patient, distinct studies            & $0.792 \pm 0.007$ & $0.841 \pm 0.013$  \\ \hline
Same patient, all studies (size controlled)           & $\mathbf{0.848 \pm 0.009}$ & $\mathbf{0.863 \pm 0.010}$  \\ \hline
\end{tabular}}
\end{table}

\subsubsection{All lateralities v.s. distinct lateralities v.s. same laterality}
First, we hypothesize that the drop in performance from the all lateralities to the same laterality strategy could be due to $\mathcal S_{\text{same study, same laterality}}(\bm{x})$ having smaller size. To test this, we carry out an experiment in which $\mathcal S_{\text{same study, all lateralities}}(\bm{x})$ is constrained by $\left| \mathcal S_{\text{same study, same laterality}}(\bm{x})\right|$, the number of images from the same study and has the same laterality as $\bm{x}$. 



\begin{table}[htbp]
\floatconts{tab:control_laterality_1}
{\caption{Experiments with all lateralities where we control the size of $S_{\text{same study, all lateralities}}$ show that the size of $\mathcal S_c(\bm x)$ affects downstream performance.}}
{\begin{tabular}{|l|r|r|}
\hline
\textbf{Criteria $c$ for creating} $\mathcal S_c(\bm{x})$             & \textbf{Linear}   & \textbf{End-to-end}   \\ \hline \hline
\begin{tabular}[c]{@{}l@{}}Same patient, same study, same laterality\end{tabular} & $0.862 \pm 0.004$ & $0.894 \pm 0.013$        \\ \hline
Same patient, same study, all lateralities (size controlled) & $0.860 \pm 0.004$                    & $0.899 \pm 0.011$                     \\ \hline
Same patient, same study, all lateralities (no control) & $\mathbf{0.876 \pm 0.013}$                    & $\mathbf{0.902 \pm 0.007}$             \\ \hline
\end{tabular}}
\end{table}

\begin{table}[htbp]
\floatconts
{tab:control_laterality_2}
{\caption{Experiments to compare same v.s. distinct lateralities with size restriction on $\mathcal S_c(\bm x)$ shows no significant difference.}}
{\begin{tabular}{|l|r|r|}
\hline
\textbf{Criteria $c$ for creating} $\mathcal S_c(\bm{x})$             & \textbf{Linear}   & \textbf{End-to-end}   \\ \hline \hline
Same patient, same study, same laterality                                               & $0.856 \pm 0.016$          &    $0.878 \pm 0.015$                       \\ \hline
\begin{tabular}[c]{@{}l@{}}Same patient, same study, distinct lateralities\end{tabular} & $\mathbf{0.866 \pm 0.015}$           &  $\mathbf{0.882 \pm 0.017}$                         \\ \hline
\end{tabular}}
\end{table}
Our second hypothesis is that mutual information in images with different lateralities is lower, which benefits retaining only information important to the downstream task, as shown in \citet{goodview}. We test this by training two models on images that include at least one counterpart from the other laterality. We pretrain one model with $\mathcal S_{\text{same study, same laterality}}(\bm{x})$ containing only images with the same laterality as $\bm{x}$, and the other model with $\mathcal S_{\text{same study, distinct lateralities}}(\bm{x})$ containing only images with different laterality from $\bm{x}$. To prevent the effect of different sizes of $\mathcal S_c(\bm{x})$, we force that $\left| \mathcal S_{\text{same study, same laterality}}(\bm{x}) \right| = \left| \mathcal S_{\text{same study, distinct lateralities}}(\bm{x}) \right|$ via random subset pre-selection.

\paragraph{Results. } Table \ref{tab:control_laterality_1} shows that once we control for the size of $\mathcal S_c(\bm x)$, there is no significant difference between using images from the same laterality (AUC = 0.862) or from all lateralities (AUC = 0.860). However, the model pretraining with all images from all lateralities  achieves much larger downstream AUC of 0.876. Thus, it supports our first hypothesis that the size of $\mathcal S_c(\bm{x})$ influences pretrained representation quality. Table \ref{tab:control_laterality_2} shows that once we control for the size of $\mathcal S_c(\bm x)$, the model pretrained with images from different lateralities only gain $0.010$ AUC in linear fine-tuning performance and a non-significant $0.004$ in end-to-end performance. This experiment shows that the effect of mutual information from different lateralities on pretrained representation quality is less pronounced.

\subsection{Negative pair selection}
We explore strategies using metadata in the CheXpert dataset to define negative pairs. 
Similar to our method of defining positive pairs, we take advantage of metadata available in the dataset to select the negative pairs. However, unlike positive pair selection, where only a single pair is required for each image, an image has to pair with the entire queue to select negative pairs. This property makes selecting negative pairs from the same patient as done in selecting positive pairs not suitable because only a small number of images are available for a patient.  We instead use a more general property -- laterality -- across the patients to define negative pairs to retain sufficient negative pairs in the loss function (\ref{loss}).  
 Similarly, other metadata such as age and sex may be exploited for the same purpose.
 
The default negative pair selection strategy is to select all keys from the queue that are not views of the query image. However, we hypothesize that negative pairs with the same laterality are ``hard'' negative pairs that are more difficult to distinguish and provide more accurate pretrained representations for the downstream task. We describe our four strategies briefly as follows and in more detail in Appendix \ref{app}.  Our first strategy is to only select images from the queue with the same laterality as the query to create negative pairs. Our second strategy is to reweight the negative logits based on laterality so in effect queries with each laterality (frontal and lateral) equally contribute to the loss and the queue size remains fixed as in the original MoCo approach. Following a similar idea in \citet{kalantidis2020hard}, our third strategy is to sample a portion of negative pairs with the same laterality for each query and append them to the queue for loss computation. Our fourth strategy is to create synthetic negatives for additional hard negative pairs.  Unlike \citet{kalantidis2020hard}, we do not determine hardness of negative pairs based on similarities of representations. Instead, we use existing metadata (image laterality) to approximate hardness of an negative pair. 
We evaluate the performance of each of these negative pair strategies combined with the positive pair strategy of ``same patient, same study, all lateralities''.

\paragraph{Results.} Results are given in Table \ref{tab:negative_results}.
The default negative pair selection strategy (AUC = 0.876) is not outperformed by any of the metadata-exploiting negative pair selection strategies including same laterality only (AUC = 0.872), same laterality reweighted (AUC = 0.864 ), same laterality appended (AUC = 0.875) and same laterality synthetic (AUC = 0.870). 
Thus, our exploratory analysis does not indicate sufficient evidence for performance improvement using strategies that incorporate metadata, but further experiments with other metadata sources may be required to further understand this relationship.

\begin{table}[htbp]
\floatconts
{tab:negative_results}
{\caption{Experiments with the default negative pair definition (different images) and various negative pair selection strategies.} }
{\begin{tabular}{|l|r|}
\hline
\textbf{Negative Pairs Strategy}             & \textbf{Linear}   \\ \hline \hline
Default &  $\mathbf{0.876 \pm 0.013}$      \\ \hline
Same Laterality only & $0.872 \pm 0.011$                   \\ \hline
Same Laterality (reweighted) & $0.864 \pm 0.006$  \\ \hline
Same Laterality (appended) & $0.875 \pm 0.006$  \\ \hline
Same Laterality (synthetic) & $0.870 \pm 0.004$  \\ \hline
\end{tabular}}
\end{table}
\vspace{-12pt}



\section{Discussion} 


We introduce MedAug, a method to use patient metadata to select positive pairs for contrastive learning, and demonstrate the utility of this method on a chest X-ray interpretation task. 

\textit{Can we improve performance by leveraging  metadata to choose positive pairs?} Yes. Our best pretrained strategy with multiple images from the same patient and same study obtains an increase of 3.4\% in linear fine-tuning AUC in comparison to the instance discrimination approach implemented in \citet{sowrirajan}. A similar result has been shown by \citet{clocs} for ECG signal interpretation.  \citet{azizi} also found improvement in dermatology classification when applying a second contrastive pretraining stage where strictly distinct images from the same patient are selected as positive pairs. 



Unlike previous work, our empirical analysis on using images from all studies and distinct studies shows that simply choosing images from the same patient may hurt downstream performance. We show that using appropriate metadata such as study number to select positive pairs that share underlying disease information is needed to obtain the best representation for the downstream task of disease classification. 
For future studies, it is of interest to experiment with other metadata such as age group, medical history, etc. and how they can inform on tasks other than disease classification.



Our analysis using different criteria on laterality shows that the number of images selected to form positive pairs plays an important role, while the effect of mutual information is less clear. Given time and resources, it would be informative to experiment with how the maximum number of distinct images chosen per query affect downstream performance. 

\textit{Can we improve performance by leveraging metadata to choose hard negative pairs?} Not necessarily. We perform an exploratory analysis of strategies to leverage patient metadata to select negative pairs, and do not find them to outperform the baseline. 

In closing, our work demonstrates the potential benefits of incorporating patient metadata into self-supervised contrastive learning for medical images, and can be extended to a broader set of tasks \citep{rajpurkar2020appendixnet, uyumazturk2019deep}. 

\paragraph{Limitations}

 In the CheXpert dataset, any two images from the same patient and study will always have the same set of ground truth pathology labels. For medical image datasets with different constraints regarding patient metadata, it remains future work to determine the positive pair selection strategies that are clinically relevant and produce good pretrained representations. Our approach is not applicable to datasets lacking patient metadata altogether. For datasets with limited data per patient, future work could be to cluster data using the images and available metadata into larger groups, and define positive pairs based on cluster assignments. While the results show that our metadata-based contrastive learning methods are generalizable across all CheXpert competition tasks, it remains future work to assess performance in other datasets.

 Furthermore, our strategies for negative pair selection did not improve pretrained representations. Our strategies leveraged information regarding image laterality. However, future work is required to whether negative pair selection strategies using other metadata such as image view (anteroposterior or posteroanterior), patient age or patient sex, or strategies using similarity metrics can improve negative pair selection. 
 



\bibliography{sample}

\begin{thebibliography}{19}
\providecommand{\natexlab}[1]{#1}
\providecommand{\url}[1]{\texttt{#1}}
\expandafter\ifx\csname urlstyle\endcsname\relax
  \providecommand{\doi}[1]{doi: #1}\else
  \providecommand{\doi}{doi: \begingroup \urlstyle{rm}\Url}\fi

\bibitem[Azizi et~al.(2021)Azizi, Mustafa, Ryan, Beaver, Freyberg, Deaton, Loh,
  Karthikesalingam, Kornblith, Chen, et~al.]{azizi}
Shekoofeh Azizi, Basil Mustafa, Fiona Ryan, Zachary Beaver, Jan Freyberg,
  Jonathan Deaton, Aaron Loh, Alan Karthikesalingam, Simon Kornblith, Ting
  Chen, et~al.
\newblock Big self-supervised models advance medical image classification.
\newblock \emph{arXiv preprint arXiv:2101.05224}, 2021.

\bibitem[Chaitanya et~al.(2020)Chaitanya, Erdil, Karani, and
  Konukoglu]{localglobalcontrastive}
Krishna Chaitanya, Ertunc Erdil, Neerav Karani, and Ender Konukoglu.
\newblock Contrastive learning of global and local features for medical image
  segmentation with limited annotations.
\newblock \emph{arXiv preprint arXiv:2006.10511}, 2020.

\bibitem[Chen et~al.(2020{\natexlab{a}})Chen, Kornblith, Swersky, Norouzi, and
  Hinton]{simclrv2}
Ting Chen, Simon Kornblith, Kevin Swersky, Mohammad Norouzi, and Geoffrey
  Hinton.
\newblock Big self-supervised models are strong semi-supervised learners.
\newblock \emph{arXiv preprint arXiv:2006.10029}, 2020{\natexlab{a}}.

\bibitem[Chen et~al.(2020{\natexlab{b}})Chen, Fan, Girshick, and He]{mocov2}
Xinlei Chen, Haoqi Fan, Ross Girshick, and Kaiming He.
\newblock Improved baselines with momentum contrastive learning.
\newblock \emph{arXiv preprint arXiv:2003.04297}, 2020{\natexlab{b}}.

\bibitem[Hjelm et~al.(2018)Hjelm, Fedorov, Lavoie-Marchildon, Grewal, Bachman,
  Trischler, and Bengio]{deepinfomax}
R~Devon Hjelm, Alex Fedorov, Samuel Lavoie-Marchildon, Karan Grewal, Phil
  Bachman, Adam Trischler, and Yoshua Bengio.
\newblock Learning deep representations by mutual information estimation and
  maximization.
\newblock \emph{arXiv preprint arXiv:1808.06670}, 2018.

\bibitem[Irvin et~al.(2019)Irvin, Rajpurkar, Ko, Yu, Ciurea-Ilcus, Chute,
  Marklund, Haghgoo, Ball, Shpanskaya, et~al.]{chexpert}
Jeremy Irvin, Pranav Rajpurkar, Michael Ko, Yifan Yu, Silviana Ciurea-Ilcus,
  Chris Chute, Henrik Marklund, Behzad Haghgoo, Robyn Ball, Katie Shpanskaya,
  et~al.
\newblock Chexpert: A large chest radiograph dataset with uncertainty labels
  and expert comparison.
\newblock In \emph{Proceedings of the AAAI Conference on Artificial
  Intelligence}, volume~33, pages 590--597, 2019.

\bibitem[Johnson et~al.(2017)Johnson, Douze, and J{\'e}gou]{JDH17}
Jeff Johnson, Matthijs Douze, and Herv{\'e} J{\'e}gou.
\newblock Billion-scale similarity search with gpus.
\newblock \emph{arXiv preprint arXiv:1702.08734}, 2017.

\bibitem[Kalantidis et~al.(2020)Kalantidis, Sariyildiz, Pion, Weinzaepfel, and
  Larlus]{kalantidis2020hard}
Yannis Kalantidis, Mert~Bulent Sariyildiz, Noe Pion, Philippe Weinzaepfel, and
  Diane Larlus.
\newblock Hard negative mixing for contrastive learning, 2020.

\bibitem[Kiyasseh et~al.(2020)Kiyasseh, Zhu, and Clifton]{clocs}
Dani Kiyasseh, Tingting Zhu, and David~A Clifton.
\newblock Clocs: Contrastive learning of cardiac signals.
\newblock \emph{arXiv preprint arXiv:2005.13249}, 2020.

\bibitem[Oord et~al.(2018)Oord, Li, and Vinyals]{cpc}
Aaron van~den Oord, Yazhe Li, and Oriol Vinyals.
\newblock Representation learning with contrastive predictive coding.
\newblock \emph{arXiv preprint arXiv:1807.03748}, 2018.

\bibitem[Rajpurkar et~al.(2020)Rajpurkar, Park, Irvin, Chute, Bereket,
  Mastrodicasa, Langlotz, Lungren, Ng, and Patel]{rajpurkar2020appendixnet}
Pranav Rajpurkar, Allison Park, Jeremy Irvin, Chris Chute, Michael Bereket,
  Domenico Mastrodicasa, Curtis~P Langlotz, Matthew~P Lungren, Andrew~Y Ng, and
  Bhavik~N Patel.
\newblock Appendixnet: Deep learning for diagnosis of appendicitis from a small
  dataset of ct exams using video pretraining.
\newblock \emph{Scientific reports}, 10\penalty0 (1):\penalty0 1--7, 2020.

\bibitem[Sowrirajan et~al.(2020)Sowrirajan, Yang, Ng, and
  Rajpurkar]{sowrirajan}
Hari Sowrirajan, Jingbo Yang, Andrew~Y Ng, and Pranav Rajpurkar.
\newblock Moco pretraining improves representation and transferability of chest
  x-ray models.
\newblock \emph{arXiv preprint arXiv:2010.05352}, 2020.

\bibitem[Sriram et~al.(2021)Sriram, Muckley, Sinha, Shamout, Pineau, Geras,
  Azour, Aphinyanaphongs, Yakubova, and Moore]{sriram2021covid}
A~Sriram, M~Muckley, K~Sinha, F~Shamout, J~Pineau, KJ~Geras, L~Azour,
  Y~Aphinyanaphongs, N~Yakubova, and W~Moore.
\newblock Covid-19 prognosis via self-supervised representation learning and
  multi-image prediction.
\newblock 2021.

\bibitem[Tamkin et~al.(2020)Tamkin, Wu, and Goodman]{tamkin2020viewmaker}
Alex Tamkin, Mike Wu, and Noah Goodman.
\newblock Viewmaker networks: Learning views for unsupervised representation
  learning, 2020.

\bibitem[Tian et~al.(2020)Tian, Sun, Poole, Krishnan, Schmid, and
  Isola]{goodview}
Yonglong Tian, Chen Sun, Ben Poole, Dilip Krishnan, Cordelia Schmid, and
  Phillip Isola.
\newblock What makes for good views for contrastive learning.
\newblock \emph{arXiv preprint arXiv:2005.10243}, 2020.

\bibitem[Uyumazturk et~al.(2019)Uyumazturk, Kiani, Rajpurkar, Wang, Ball, Gao,
  Yu, Jones, Langlotz, Martin, Berry, Ozawa, Hazard, Brown, Chen, Wood, Allard,
  Ylagan, Ng, and Shen]{uyumazturk2019deep}
Bora Uyumazturk, Amirhossein Kiani, Pranav Rajpurkar, Alex Wang, Robyn~L. Ball,
  Rebecca Gao, Yifan Yu, Erik Jones, Curtis~P. Langlotz, Brock Martin,
  Gerald~J. Berry, Michael~G. Ozawa, Florette~K. Hazard, Ryanne~A. Brown,
  Simon~B. Chen, Mona Wood, Libby~S. Allard, Lourdes Ylagan, Andrew~Y. Ng, and
  Jeanne Shen.
\newblock Deep learning for the digital pathologic diagnosis of
  cholangiocarcinoma and hepatocellular carcinoma: Evaluating the impact of a
  web-based diagnostic assistant, 2019.

\bibitem[Wu et~al.(2018{\natexlab{a}})Wu, Xiong, Yu, and
  Lin]{wu2018unsupervised}
Zhirong Wu, Yuanjun Xiong, Stella Yu, and Dahua Lin.
\newblock Unsupervised feature learning via non-parametric instance-level
  discrimination, 2018{\natexlab{a}}.

\bibitem[Wu et~al.(2018{\natexlab{b}})Wu, Xiong, Yu, and Lin]{instancedisc}
Zhirong Wu, Yuanjun Xiong, Stella~X Yu, and Dahua Lin.
\newblock Unsupervised feature learning via non-parametric instance
  discrimination.
\newblock In \emph{Proceedings of the IEEE Conference on Computer Vision and
  Pattern Recognition}, pages 3733--3742, 2018{\natexlab{b}}.

\bibitem[Zhang et~al.(2020)Zhang, Jiang, Miura, Manning, and
  Langlotz]{textimagecontrastive}
Yuhao Zhang, Hang Jiang, Yasuhide Miura, Christopher~D Manning, and Curtis~P
  Langlotz.
\newblock Contrastive learning of medical visual representations from paired
  images and text.
\newblock \emph{arXiv preprint arXiv:2010.00747}, 2020.

\end{thebibliography}

\appendix

\section{Repetition Across Tasks}\label{section:new}
We repeated each linear fine-tuning experiment for the other 4 CheXpert competition classification tasks: Atelectasis, Cardiomegaly, Consolidation, and Edema. The results are given in Table \ref {tab:new}. For all 5 CheXpert competition classification tasks (see Table \ref{tab:positive} for Pleural Effusion), we found that our best metadata-based approach to contrastive learning, which was the “same patient, same study” positive pair selection criterion, resulted in higher downstream classification performance than did the ImageNet baseline and MoCo v2 baselines.

\begin{table}[htbp!]
\scriptsize
\floatconts
{tab:new}
{\caption{Linear fine-tuning AUCs for baseline methods and positive pair selection criteria repeated across the tasks of Atelectasis, Cardiomegaly, Consolidation, and Edema classification.}}
{\begin{tabular}{|l|c|c|c|c|}
\hline
\textbf{Baseline models} & \textbf{Atelectasis} & \textbf{Cardiomegaly} & \textbf{Consolidation} & \textbf{Edema}
\\
\hline \hline 
ImageNet baseline & 0.612 & 0.634 & 0.615 & 0.810 \\
\hline
MoCo v2 baseline \citep{sowrirajan}                                                               & 0.671 & 0.735 & 0.699 &  0.847      \\ \hline
MoCo v2 baseline with random crop scale                                               & 0.708 & 0.730 & 0.693 &  0.858        \\ \hline 
\hline \textbf{Criteria $c$ for creating} $\mathcal S_c(\bm{x})$ & \textbf{Atelectasis} & \textbf{Cardiomegaly} & \textbf{Consolidation} & \textbf{Edema}           \\ \hline \hline
\begin{tabular}[c]{@{}l@{}}Same patient, same study, same laterality\end{tabular} & 0.738 & 0.718 & 0.672 & 0.866       \\ \hline
Same patient, same study, distinct lateralities   & 0.712 & 0.800 & 0.703 & 0.883          \\ \hline
\begin{tabular}[c]{@{}l@{}}Same patient, same study \end{tabular}    & 0.762 & 0.785 & 0.721 & 0.889 \\ \hline
\begin{tabular}[c]{@{}l@{}}Same patient, all studies \end{tabular}                   & 0.727 & 0.786 & 0.805 &  0.879     \\ \hline
\begin{tabular}[c]{@{}l@{}}Same patient, distinct studies \end{tabular}                   & 0.676 & 0.757 & 0.760 &    0.846       \\ \hline
\begin{tabular}[c]{@{}l@{}}Same patient, same study with random crop scale   \end{tabular}                           & 0.721 & 0.779 & 0.801 &   0.866      \\ \hline
\end{tabular}}
\end{table}

\section{Negative Pairs}\label{app}
Following the loss function in equation (1), we denote the exponential sum of the negative pairs by $G$
\begin{align}\mathcal{L}(\bm{x})=-\log\frac{\exp[g(\bm{\tilde{x}}_{1})\cdot g(\bm{\tilde{x}}_{2})]}{\exp[g(\bm{\tilde{x}}_{1})\cdot g(\bm{\tilde{x}}_{2})]+G\left(\bm{\tilde{x}}_{1},\bm{z}_{i}\right)}.
\end{align}
where
\begin{align}G\left(\bm{\tilde{x}}_{1},\bm{z}_{i}\right)=\sum_{\bm{z}_{i}\in Q}\exp[g(\bm{\tilde{x}}_{1}))\cdot g(\bm{z}_{i}))]
\end{align} We follow the MoCo setup and denote $Q$ as the queue. Let $\mathcal{S}(\bm{x})$ be the set
of image representations in $Q$ that have the same laterality as $\bm{x}$. We use the symbol $\mathbin\Vert$ to denote list concatenation. We describe each of our negative pair selection strategies as follows:

\begin{enumerate}
\item(Same laterality only) For each query, we select keys in the queue that have the same laterality as the query. Specifically, we replace $G$ in equation (4) by $G^l$

\begin{align}
G^l\left(\bm{\tilde{x}}_{1},\bm{z}_{i}\right)=\sum_{ \bm{z}_{i}\in \mathcal S(\bm{x})}\exp[g(\bm{\tilde{x}}_{1}))\cdot g(\bm{z}_{i}))]
\end{align}

\item(Same laterality reweighted) The first strategy excluded the keys in the queue that have different laterality from the query. Here we set a target hard negative weight and reweight each $exp$ term to achieve the target weight.  Let

\begin{align}
G^w\left(\bm{\tilde{x}}_{1},\bm{z}_{i}\right)=\sum_{ \bm{z}_{i}\in \mathcal S(\bm{x})}w_{i}^{s}\exp[g(\bm{\tilde{x}}_{1}))\cdot g(\bm{z}_{i}))]+\sum_{ \bm{z}_{i}\in \mathcal S\left(x\right)^{c}}w_{i}^{d}\exp[g(\bm{\tilde{x}}_{1}))\cdot g(\bm{z}_{i}))]
\end{align}

where $t$ is the target hard negative weight  and $r=\frac{\vert \mathcal S(\bm{x})\vert}{\vert Q \vert}$ is the proportion of the negative keys in the queue that have the same laterality as $\bm{x}$.  Then $w_{i}^{d}=\frac{1-t}{1-r}$ and
$w_{i}^{s}=\frac{t}{r}$ for all $i$.  
In our experiments, we set $t=0.1$ to allocate 90\% of the weight to hard negatives. This allows us to include all negative pairs in the contrastive loss but place emphasis on hard negative pairs with the same laterality. 

\item(Same laterality appended)
For each query, we select a random sample of the keys that have the same laterality and append them to the existing queue

\[
Q= [z_1,z_2,...,z_K]
\]
where $K$ is the queue size. Let 
\begin{align}
A = \left\{ z_{i_{1}},z_{i_{2}},...,z_{i_{m}}\right\} \subset S\left(x\right)
\end{align}
be the random sample of keys with the same laterality as the query. The new queue is
\[
Q^{a}=Q \mathbin\Vert A
\]
and 
\[
G^{a}\left(\bm{\tilde{x}}_{1},\bm{z}_{i}\right)=\sum_{\bm{z}_{i}\in Q^{a}}\exp[g(\bm{\tilde{x}}_{1}))\cdot g(\bm{z}_{i}))]
\]
replaces $G$ in equation (4).

\item(Same laterality synthetic)
For each query, in addition to appending samples of the keys from $S(\bm{x})$, we use the samples to generate synthetic keys and append them to the queue.  We randomly sample $m$ pairs $\left(\bm{s}_{i},\bm{s}_{j}\right)\in A =\left\{ \bm{z}_{i_{1}},\bm{z}_{i_{2}},...,\bm{z}_{i_{m}}\right\} $ and call this set of pairs $B$. 

For each pair $\left(\bm{s}_{i},\bm{s}_{j}\right)\in B$, we uniformly sample a number $u$ between 0 and 1 and let
\[h=u\cdot \bm{s}_{i}+(1-u)\cdot \bm{s}_{j}\] A synthetic image
representation is defined as the normalized vector $\frac{h}{\|h\|}$.
Let $H$ be the set of these $m$ synthetic image representations and 
\[
Q^{h}=Q \mathbin\Vert B \mathbin\Vert H
\] is the new queue. $G$ in equation (1) is replaced by
\[
G^{h}\left(\bm{\tilde{x}}_{1},\bm{z}_{i}\right)=\sum_{z_{i}\in Q^{h}}\exp[g(\bm{\tilde{x}}_{1}))\cdot g(\bm{z}_{i}))]
\]
Note that unlike \citet{kalantidis2020hard}, we only construct synthetic images once.
\end{enumerate}

\end{document}